\documentclass[prl,twocolumn,preprintnumbers,amsmath,amssymb,showpacs]{revtex4}
\usepackage{epsfig}
\usepackage[bookmarks = false,plainpages = false,hypertexnames = false]{hyperref}  
\usepackage{color}
\usepackage{subfigure}
\usepackage{epstopdf}

\begin{document}
\allowdisplaybreaks
\setlength{\voffset}{1.0cm}
\title{Quantum anomaly, universal relations, and breathing mode of a two-dimensional Fermi gas}
\author{Johannes Hofmann}
\email{j.b.hofmann@damtp.cam.ac.uk}
\affiliation{Department of Applied Mathematics and Theoretical Physics, University of Cambridge, Centre for Mathematical Sciences,
Cambridge CB3 0WA, United Kingdom}
\date{\today}
\begin{abstract}
In this Letter, we show that the classical $SO(2,1)$ symmetry of a harmonically trapped Fermi gas in two dimensions is broken by quantum effects. The anomalous correction to the symmetry algebra is given by a two-body operator that is well known as the \emph{contact}. Taking into account this modification, we are able to derive the virial theorem for the system and a universal relation for the pressure of a homogeneous gas. The existence of an undamped breathing mode is associated with the classical symmetry. We provide an estimate for the anomalous frequency shift of this oscillation at zero temperature and compare the result with a recent experiment by [E. Vogt {\sl et al.}, Phys. Rev. Lett. {\bf 108}, 070404 (2012)]. Discrepancies are attributed to finite temperature effects.
\end{abstract}
\maketitle

The symmetry of a classical theory can be broken in the corresponding quantum theory. This is known as a quantum anomaly. Probably the best-known example of an anomaly is the conformal anomaly: While the classical theory only contains dimensionless bare parameters, the quantization process forces us to introduce a regulator scale so that divergent terms are canceled. The bare quantities and the regulator conspire in such a way that the renormalized quantities are dimensionful. This startling phenomenon is called dimensional transmutation. It is ubiquitous in quantum field theory, examples include QED, QCD, and the Gross--Neveu model.

A quantum anomaly occurs for a two-dimensional (2D) gas of two fermion species as well. In this case the anomaly breaks classical scale invariance, and is therefore referred to as a scale or conformal anomaly. The Hamiltonian of the system can be written as
\begin{align}
 H &= \int d^2x \, \left[\psi_\sigma^\dagger \frac{- \nabla^2}{2 m} \psi_\sigma({\bf x}) + \frac{\lambda}{m} \psi_\uparrow^\dagger \psi_\downarrow^\dagger \psi_\downarrow^{} \psi_\uparrow^{}({\bf x})\right] . \label{eq:1} 
\end{align}
We can read off the engineering dimension of the fields and the coupling~\cite{ref:t7}: The fermion fields $\psi_\sigma({\bf x})$ (with $\sigma = \uparrow, \downarrow$) have a dimension of momentum whereas the bare coupling constant $\lambda$ is dimensionless. Since we work with a local Hamiltonian, the quantum field theory contains divergences which must be removed by a renormalization of $\lambda$. This is done in such a way that the leading-order term in the effective range expansion of the quantum-mechanical $s$-wave scattering phase shift is reproduced. Hence, the Hamiltonian~(\ref{eq:1}) is referred to as the zero-range model. The renormalized dimensionful parameter is the bound state energy $E_b$. With a hard momentum space cutoff $\Lambda$, $E_b$ is related to $\lambda$ and the regulator scale $\Lambda$ by~\cite{ref:t6,ref:25}
\begin{align}
 E_b &= - \frac{4}{m a_{\rm 2D}^2 e^{2 \gamma_E}} = - \frac{\Lambda^2}{m} e^{4\pi/\lambda} . \label{eq:2}
\end{align}
We set $\hbar \equiv 1$. $a_{\rm 2D}$ is the scattering length and $\gamma_E$ is Euler's constant. As the cutoff diverges, $\Lambda \rightarrow \infty$, the bare coupling $\lambda(\Lambda) \nobreak \rightarrow \nobreak - \nobreak 0$ so that $E_b$ is kept fixed.
 
Experimentally, a 2D Fermi gas can be created by trapping an ultracold quantum gas in a geometry that tightly confines the gas in one direction \cite{ref:1,ref:2,ref:3}. Such a system can be described by the following effective Hamiltonian:
\begin{align}
 H_{\rm osc} = H + \int d^2x \, \frac{m \omega_0^2 x^2}{2} \psi_\sigma^\dagger \psi_\sigma^{}({\bf x}) , \label{eq:3}
\end{align}
where $\omega_0$ is the azimuthal trapping frequency. Although the trapping potential explicitly breaks scale invariance, the system still exhibits a symmetry on the classical level~\cite{ref:4}. The symmetry group is $SO(2, 1)$, the Lorentz group in 2D. The generators of the corresponding algebra are constructed from the generators of scale and special conformal transformations,
\begin{align}
 D &=\int d^2x \, x_i m j_i({\bf x})~{\rm and}~C=\int d^2x \frac{m x^2}{2} \psi_\sigma^\dagger \psi_\sigma^{}({\bf x}) ,
\end{align}
as well as the Hamiltonian of the noninteracting system, Eq.~(\ref{eq:1}). $j_i = -i(\psi^\dagger \partial_i \psi - \partial_i \psi^\dagger \psi)/2m$ is the current operator. 
We can express the Hamiltonian of the trapped system~(\ref{eq:3}) as a linear combination of two generators: $H_{\rm osc} = H + \omega_0^2 C$. $D$, $C$, and $H$ obey the commutation relations \cite{ref:5,ref:6}
\begin{align}
 [D, H] = 2 i H, \ [D, C] = - 2 i C, \ {\rm and} \ [H, C] = - i D . \label{eq:5}
\end{align}
The commutators in Eq.~(\ref{eq:5}) hold on the classical level, where they are understood in terms of Poisson brackets $[A,B]_{\rm PB} \equiv \int d^2x \, (\tfrac{\delta A}{\delta^{} \psi_\sigma^{}} \tfrac{\delta B}{\delta \psi_\sigma^\dagger} - \tfrac{\delta A}{\delta^{} \psi_\sigma^\dagger} \tfrac{\delta B}{\delta \psi_\sigma})$. There are quantum corrections to those relations, and one of the main results of this Letter is to derive the anomalous correction to the commutator $[D, H_{\rm osc}]$, Eq.~(\ref{eq:13}). The operators $L_{1/2} = (L_+ \pm L_-)/4$, where $L_{\pm} = (H - \omega_0^2 C \pm i \omega_0 D)/\omega_0$, and the Hamiltonian $L_3 = H_{\rm osc}/2\omega_0$ form the $so(2,1)$ algebra $[L_1, L_2] = L_3$, $[L_2, L_3] = - L_1$, and $[L_3, L_1] = L_2$. This is an example of a spectrum generating symmetry: $L_+$ and $L_-$ are raising and lowering operators with commutators $[H_{\rm osc}, L_\pm] = \pm 2 \omega_0 L_\pm$ and $[L_+, L_-] = - 4 H_{\rm osc}/\omega_0$. If $L_\pm$ acts on an eigenstate of the Hamiltonian $H_{\rm osc}$ with energy $E$, we obtain an eigenstate with energy $E \pm 2 \omega_0$. The excitations of the ground state were identified with breathing modes~\cite{ref:4}. The work in Ref.~\cite{ref:4} has been generalized to the unitary Fermi gas in three dimensions for which the symmetry persists on the quantum level \cite{ref:5,ref:6,ref:7}.

Since the $SO(2,1)$ symmetry on the classical level is used to predict the breathing mode spectrum, the quantum anomaly should induce a shift in that spectrum. We provide an estimate for this shift in the hydrodynamic regime at zero temperature based on a recent Monte Carlo simulation of the equation of state \cite{ref:17}.

In this Letter, we show that quantum effects deform the algebra~(\ref{eq:5}), which breaks the symmetry. We find that this anomaly is closely related to a two-body operator known as the \emph{contact}, which plays an important role in the description of various properties of an interacting Fermi gas. This work provides a new interpretation of the contact in 2D. The contact operator is defined as
\begin{align}
 I &= \int d^dx \, \lambda^2 \psi_\uparrow^\dagger \psi_\downarrow^\dagger \psi_\downarrow^{} \psi_\uparrow^{}({\bf x}) , \label{eq:6}
\end{align}
where $d$ is the space dimension. As is well known, the matrix elements of $I$ are finite \cite{ref:8}. They set the magnitude of a class of exact relations that are referred to as universal or Tan relations~\cite{ref:9}. They include several thermodynamic relations as well as the asymptotic form of various correlators, such as the momentum distribution, structure factors, or radio-frequency transition rates. For a review, see Ref.~\cite{ref:10}. Universal relations for Fermi gases in 2D have been derived in Refs.~\cite{ref:11a,ref:11,ref:12,ref:13,ref:14}. Although we consider a fermionic theory in this Letter, all our results can be extended to bosonic systems.

\emph{Quantum anomaly}.---We start by deriving the form of the quantum anomaly. We show that the commutator of $D$ and $H$ receives an anomalous correction which is proportional to the contact operator $I$. This operator is not an element of the original $so(2,1)$ algebra: The algebra is deformed and the symmetry is broken. The effect of a quantum anomaly on a Bose gas at zero temperature has been studied previously in Refs.~\cite{ref:15,ref:16}.

The change of the Hamiltonian under an infinitesimal scale transformation is given by $[D, H_{\rm osc}]$. We can simplify this expression using the Euler equation
\begin{align}
 - i m [j_i({\bf x}), H_{\rm osc}]  &= - \partial_j \Pi_{ij} - x_i m \omega_0^2 \psi_\sigma^\dagger \psi_\sigma^{}({\bf x}) ,
\end{align}
where $\Pi_{ij}$ is the stress tensor of the zero-range model,
\begin{align}
\Pi_{ij}({\bf x}) &= \frac{1}{2m} \left[ \partial_i \psi_\sigma^\dagger \partial_j \psi_\sigma + \partial_j \psi_\sigma^\dagger \partial_i \psi_\sigma - \frac{\delta_{ij}}{2} \nabla^2 (\psi_\sigma^\dagger \psi_\sigma) \right] \nonumber \\
&\quad + \delta_{ij} \frac{\lambda}{m} \psi_\uparrow^\dagger \psi_\downarrow^\dagger \psi_\downarrow^{} \psi_\uparrow^{}({\bf x}) .
\end{align}
This yields, after integrating by parts,
\begin{align}
 &[D, H_{\rm osc}] = i \int d^dx \, \Pi_{ii}({\bf x}) - 2 i \omega_0^2 C \nonumber \\
&= 2 i H - 2 i \omega_0^2 C + i \frac{(d - 2) \lambda^{-1}}{m} I . \label{eq:9}
\end{align}
If the last term on the right-hand side of Eq.~(\ref{eq:9}) were zero, the theory would obey the commutation relations listed in Eq.~(\ref{eq:5}). In the following, we demonstrate that this term does not vanish and the trace of the stress tensor receives an anomalous correction. We choose a renormalization scheme in which $\lambda^{-1}$ diverges as $1/(d-2)$ which cancels the prefactor in Eq.~(\ref{eq:9}).
 
In order to determine the dependence of $\lambda$ on the regulator, it is sufficient to calculate few-body matrix elements. To this end, consider the scattering amplitude of two particles of opposite spin in the center of mass frame. The matrix element is obtained by summing a geometric series of ``ladder'' diagrams (see, for example, Ref.~\cite{ref:7}):
\begin{align}
 {\cal A}(E) &= [m/\lambda - G^E_{\rm osc}(0,0)]^{-1} . \label{eq:10}
\end{align}
$G^E_{\rm osc}({\bf r}', {\bf r})$ is the bare propagator of a particle with energy $E$ and reduced mass $m/2$. It is given by~\cite{ref:7}
\begin{align}
 &G^E_{\rm osc}(0, 0) = - \int_0^\infty dt \, e^{(E + i 0^+) t} \left(\frac{m \omega_0}{4 \pi \sinh \omega_0 t}\right)^{d/2} \nonumber \\
&= - \left(\frac{m}{4 \pi}\right)^{d/2} (2 \omega_0)^{d/2 - 1} \frac{\Gamma\bigl(1 - \frac{d}{2}\bigr) \Gamma\bigl(- \frac{E}{2 \omega_0} + \frac{d}{4}\bigr)}{\Gamma\bigl(- \frac{E}{2 \omega_0} - \frac{d}{4} + 1\bigr)} . \label{eq:11}
\end{align} 
Equation~(\ref{eq:11}) diverges for $d = 2$. In order to extract the divergent part, we analytically continue $d = 2 - \varepsilon$. To keep the coupling dimensionless, we introduce a dimensionful scale in Eq.~(\ref{eq:1}), which we identify with the inverse scattering length: $\lambda \rightarrow a_{\rm 2D}^{-\varepsilon} \lambda$. This yields
\begin{align}
  &a_{\rm 2D}^{- \varepsilon} G^E_{\rm osc}(0, 0) = \nonumber \\
& \ \frac{m}{4 \pi} \left(- \frac{2}{\varepsilon} + \ln \frac{m \omega_0 a_{\rm 2D}^2 e^{\gamma_E}}{2 \pi} + \psi_0\left(- \frac{E}{2 \omega_0} + \frac{1}{2}\right)\right) . \label{eq:12}
\end{align}
$\psi_0$ is the digamma function. We choose a modified minimal subtraction scheme with $\lambda^{-1}(\varepsilon) = (- 2/\varepsilon - \gamma_E - \ln \pi)/4\pi$. The poles of the scattering amplitude give the equation for the two-particle spectrum~\cite{ref:t2}. This gives rise to the relation
\begin{align}
 [D, H_{\rm osc}] &= 2 i H - 2 i \omega_0^2 C + \frac{i}{2 \pi m} I , \label{eq:13}
\end{align}
or, equivalently, $[D, H] = 2 i H + i I/2 \pi m$. The repeated action of $H$ and $D$ on $I$ generates two-particle operators that contain additional derivatives. More generally, the correction term can be written as $i \beta_\lambda \partial H/\partial \lambda$, where $\beta_\lambda = \lambda^2/2\pi$ is the beta function of the theory. The anomaly is due to the renormalization of the contact interaction and independent of the particular choice of the regulator.

Using an analytic result, the anomalous correction to a ``hydrodynamic commutator'' of a low-density Bose gas was determined in Ref.~\cite{ref:15}. We reproduce this case by using the ``adiabatic Tan relation'' $I = 4 \pi m a_{\rm 2D} \partial H/\partial a_{\rm 2D}$~\cite{ref:11} in Eq.~(\ref{eq:13}) (Note that $\beta_\lambda = \lambda^2/4 \pi$ for bosons).

We can repeat the analysis for an untrapped system. Here, the system is not only $SO(2,1)$ but also scale-invariant and the deformation of the algebra indicates a scale anomaly. Using the propagator of a homogeneous system, $G^E_{\rm hom}(0, 0) = - \int_0^\infty dt \, e^{(E + i 0^+) t} \left(m/4 \pi t\right)^{d/2}$, in Eq.~(\ref{eq:10}), we find:
\begin{align}
 &a_{\rm 2D}^{- \varepsilon} G^E_{\rm hom}(0, 0) = \frac{m}{4 \pi} \left(- \frac{2}{\varepsilon} + \ln \frac{m E a_{\rm 2D}^2 e^{\gamma_E}}{4 \pi} - i \pi\right) .
\end{align}
The choice for $\lambda(\varepsilon)$ remains unchanged. The trace of the stress tensor becomes
\begin{align}
\int d^dx \, \Pi_{ii} &= 2 H + \frac{1}{2 \pi m} I. \label{eq:15}
\end{align}
The conformal anomaly in the homogeneous case has been discussed in detail, both in quantum field theory~\cite{ref:19} and as a quantum-mechanical toy model~\cite{ref:t6,ref:18,ref:t3}. For the discussion of an enlarged model with an additional long-range interaction, see Ref.~\cite{ref:t1}.

\emph{Virial theorem and pressure relation}.---Although the $so(2,1)$ algebra is deformed, we can use the new commutation relation~(\ref{eq:13}) to derive one of the Tan relations, the virial theorem. The virial theorem relates the ground state energy $E_0 = \langle H_{\rm osc} \rangle$ of a many-particle system to the trapping energy $\omega_0^2 \langle C \rangle$. 

The thermal expectation value of an operator ${\cal B}$ is defined as $\langle {\cal B} \rangle = {\rm tr} \bigl\{\exp[- \beta \left(H_{\rm osc} - \mu_\sigma N_\sigma\right)] {\cal B}\bigr\}/{\rm tr} \bigl\{\exp[- \beta \left(H_{\rm osc} - \mu_\sigma N_\sigma\right)] \bigr\}$, where $\mu_\sigma$ is the chemical potential for the two spin species. Since $[H_{\rm osc}, N_\sigma] = 0$, the thermal weight factor commutes with $H_{\rm osc}$ and we have $\langle [H_{\rm osc}, {\cal B}] \rangle = 0$. Setting ${\cal B} = D$ we can express the ground state energy in the trap as follows:
\begin{align}
E_0 &= \langle H + \omega_0^2 C \rangle = 2 \omega_0^2 \langle C \rangle - \frac{1}{4 \pi m} \langle I \rangle .
\end{align}
This form of the virial theorem has been obtained before in Refs.~\cite{ref:11a,ref:13} using a different argument.

A second important thermodynamic Tan relation is the pressure relation. It holds for a homogeneous system and links the pressure $P$ and the energy density ${\cal E}$. The pressure relation follows directly from the anomalous trace of the stress tensor, Eq.~(\ref{eq:15}), which can be related to the pressure via $2 P V = \int d^2x \, \langle \Pi_{ii}({\bf x}) \rangle$. Combining this with Eq.~(\ref{eq:15}) gives the desired result:
\begin{align}
 P &= {\cal E} + \frac{{\cal I}}{4 \pi m} , \label{eq:20}
\end{align}
where ${\cal E}$ and ${\cal I}$ denote the energy and the contact density, respectively. Equation~(\ref{eq:20}) can also be derived as the virial theorem of a gas trapped in a box, but boundary terms have to be taken into account, so that $\langle [D, H] \rangle = 2 i P V$ \cite{ref:20}. Again, the contact term in Eq.~(\ref{eq:20}) is a direct consequence of the quantum anomaly. For a scale-invariant system, the pressure relation takes the well-known form $P = {\cal E}$.

\emph{Breathing mode}.---As outlined in the Introduction, the $SO(2,1)$ symmetry implies the existence of hydrodynamic breathing mode excitations, the energy levels of which are spaced by $2 \omega_0$.
A shift in the mode frequency is a manifestation of the quantum anomaly. A hydrodynamic description applies whenever the system is only slightly perturbed from thermodynamic equilibrium which is maintained by frequent collisions of the atoms. In a strongly interacting regime where $\ln k_F a_{\rm 2D} \approx 0$~\cite{ref:25, ref:17}, the scattering cross section and, thus, the collision rate is large and we expect a hydrodynamic description to be applicable. The effect of anomalies on the hydrodynamic equations has been much discussed in relativistic hydrodynamics, cf., for example, Ref.~\cite{ref:21}.

\begin{figure}[t!]
\begin{center}
\scalebox{0.8}{\includegraphics{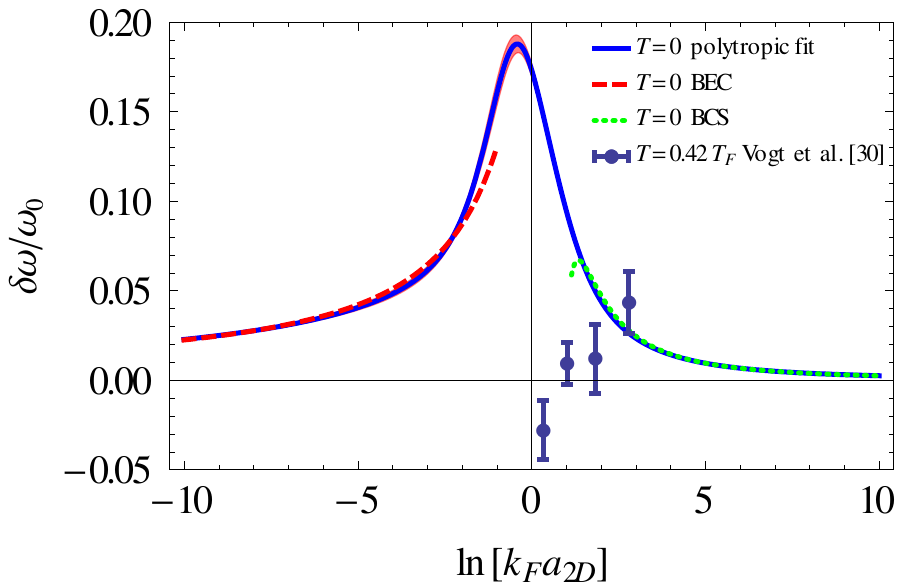}}
\caption{Anomalous frequency shift $\delta \omega$ at $T=0$ as determined from the fit to the equation of state in~\cite{ref:17}. The red error band indicates the propagated error from the Monte Carlo calculation. The plot also shows the exact results in the BCS (green dotted line) and BEC limit (red dashed line). For comparison, we include experimental values for weak breathing mode excitations at $T = 0.42 T_F$ as reported in Ref.~\cite{ref:23}. It remains an open question how thermal fluctuations affect the zero temperature results.}
\label{fig:1}
\end{center}
\end{figure}

For a scale-invariant system, dimensional analysis dictates a quadratic dependence of the pressure on the density $n$, $P \sim n^2$. If we approximate the equation of state by a polytrope, $P \sim n^{\gamma + 1}$, where $\gamma$ is the polytropic index, the linearized hydrodynamic equations can be solved (cf., for example, Ref.~\cite{ref:22}). The frequency of the breathing mode in 2D is given by $\omega^2/\omega_0^2 = 2 \gamma + 2$. We fit the numerical results for the energy per particle $E/N$ in Ref.~\cite{ref:17} to determine the polytropic index
\begin{align}
\gamma &\equiv \frac{n}{P} \frac{\partial P}{\partial n} - 1 = \frac{\xi + \xi' + \xi''/4}{\xi + \xi'/2} , \label{eq:insert}
\end{align}
where we parametrize $E/N = \tfrac{k_F^2}{4 m} \xi(\ln k_F a_{\rm 2D})$ with $k_F = \sqrt{2 \pi n}$ and define $\xi' \equiv \tfrac{\partial \xi}{\partial \ln k_F a_{\rm 2D}}$. In Eq.~(\ref{eq:insert}), the bound state part of $E/N$ does not contribute. The result of this calculation is shown in Fig.~\ref{fig:1}. The effect of the anomaly should be significant as we enter the strongly interacting regime, $\ln k_F a_{\rm 2D} \approx 0$. In this limit, we predict a shift of the order of $10 \%$ of the classical value.

In a recent experiment by E. Vogt et al., the breathing mode frequency has been measured at much higher temperature, $T/T_F \approx 0.4$ \cite{ref:23}. The authors do not observe a significant shift from the classical value and obtain a result that is roughly consistent with the mean field prediction. This suggests that at finite temperature the effect of the anomaly is washed out by thermal fluctuations. It would be very interesting to determine the equation of state at finite temperature to address this question further. (We checked that the theoretical results in Fig.~\ref{fig:1} are not significantly distorted for a slightly anisotropic trap and that they are robust under a variation of the numerical data in~\cite{ref:17}.) The experimental results in Ref.~\cite{ref:23} have been addressed in Refs.~\cite{ref:27,ref:28}, in which the damping of collective modes is analyzed using kinetic theory. 

For large values of $|\ln k_F a_{\rm 2D}|$, the energy per particle can be stated in closed analytical form. We can use this to derive analytical expressions for the frequency shifts. 
In the limit $\ln k_F a_{\rm 2D} \gg 0$, the system forms a Bardeen-Cooper-Schrieffer (BCS) superfluid and the frequency shift is
\begin{align}
 \frac{\delta \omega}{\omega_0} &= \frac{1}{4 \eta^2} - \frac{\kappa}{2 \eta^3} + {\cal O}\left(\eta^{-4}\right), \label{eq:21}
\end{align}
where $\eta = \ln k_F a_{\rm 2D}$ and $\kappa = 0.06 \pm 0.02$ has been determined in Ref.~\cite{ref:17}. Equation~(\ref{eq:21}) is indicated by the green dotted line in Fig.~\ref{fig:1}. In the Bose-Einstein condensate (BEC) limit, the system can be described as a gas of bosons with an effective dimer scattering length $a_d \approx 0.55 a_{\rm 2D}$ \cite{ref:17,ref:24}. We obtain the anomalous frequency shift
\begin{align}
 \frac{\delta \omega}{\omega_0} &= - \frac{1}{4 \eta} + {\cal O}(\eta^{-2} \ln \eta^2) , \label{eq:22}
\end{align}
which is indicated by the red dashed line in Fig.~\ref{fig:1}.

Equation~(\ref{eq:22}) holds for a Bose gas as well, where $a_d$ has to be replaced by the 2D scattering length $a_{\rm 2D}$. Following the notation in Ref.~\cite{ref:15}, we relate $a_{\rm 2D}$ to the 3D scattering length $a_{\rm 3D}$ via $a_{\rm 2D}^2 n = \sigma e^{-1/\epsilon}/\pi e^{2 \gamma_E + 1}$, where $\sigma = \pi e^{2 \gamma_E + 1} (C_{\rm 2D})^2 n \tilde{a}_z^2$, $C_{\rm 2D} \approx 1.47$ and $\epsilon = a_{\rm 3D}/\sqrt{\pi} \tilde{a}_z$, and $\tilde{a}_z = \sqrt{2/ m \omega_z}$ is the harmonic oscillator length \cite{ref:25}. If we assume $\epsilon \ll \min(1, 1/|\ln \sigma|)$, the leading-order term in Eq.~(\ref{eq:22}) reproduces the result for the anomalous frequency shift obtained by Olshanii et al., $\delta \equiv \delta \omega/2\omega_0 = a_{\rm 3D}/4\sqrt{\pi}\tilde{a}_z$ \cite{ref:15}.

It is instructive to compare the present analysis to the Fermi gas in one and three dimensions (1D/3D). There, the commutator relations between the dilatation operator and free Hamiltonian are
\begin{align}
{\rm 1D}: \quad [D, H] &= 2 i H + \frac{i a_{\rm 1D}}{2 m} I \quad {\rm and} \\
 {\rm 3D}: \quad [D, H] &= 2 i H + \frac{i a_{\rm 3D}^{-1}}{4 \pi m} I .
\end{align}
As for the 2D case, we can use the commutator relations to derive the virial theorem and the pressure relation in 1D~\cite{ref:11a,ref:26a} and in 3D~\cite{ref:8,ref:9,ref:11a}. Moreover, the symmetry is broken explicitly at finite scattering length, and there is a parameter --- the (inverse) scattering length $a_{\rm 1D}$ and $a_{\rm 3D}^{-1}$ --- that sets the strength of the breaking. For small parameter values, we can treat the correction as a small perturbation. This has been exploited to calculate the shift in hydrodynamic mode frequencies close to the unitary limit in 3D \cite{ref:26,ref:t8}.

In conclusion, we studied the effect of quantum fluctuations on the symmetry properties of a 2D Fermi gas in a harmonic trap. We showed that there is a quantum anomaly, i.e., that the Pitaevskii-Rosch symmetry which exists on a classical level is not a symmetry of the quantum system. The anomaly manifests itself in a deformation of the Lie algebra associated with the symmetry group. We derived that the anomalous operator appearing in the commutator relations is the contact operator and used this result to present a field-theoretical derivation of two thermodynamic Tan relations, the virial theorem and the pressure relation. We extracted the anomalous frequency shift of the breathing mode from Monte Carlo simulations at $T = 0$ and compared the result to recent measurements of the mode frequency at finite temperature. The findings of this Letter underline the subtle role that is played by the contact operator in the physics of interacting quantum gases. We can interpret it as an anomalous operator that is introduced by quantum fluctuations.

\emph{Acknowledgments} I thank M. Wingate for useful discussions and comments. I thank M. K\"ohl, M. Koschorreck and E. Vogt for discussions and for providing the data for Fig.~\ref{fig:1}.~I am supported by CHESS, DAAD, STFC, St.~John's College, Cambridge, and by the Studienstiftung des deutschen Volkes.

\end{document}